\documentstyle[prb,aps,epsf]{revtex} \def\narrowtext{} \tighten \twocolumn
\input epsf.sty
\begin{document}
\draft

\title{Momentum Distribution Curves in the Superconducting State}
\author{M. R. Norman,$^1$ M. Eschrig,$^1$
        A. Kaminski,$^2$ and J. C. Campuzano$^{1,2}$}
\address{         
         (1) Materials Science Division, Argonne National Laboratory,
             Argonne, IL 60439 \\
         (2) Department of Physics, University of Illinois at Chicago,
             Chicago, IL 60607\\
         }

\address{%
\begin{minipage}[t]{6.0in}
\begin{abstract}
We demonstrate that the E vs.~{\bf k} dispersion in the superconducting 
state extracted from momentum distribution curves differs qualitatively
from the traditional dispersion extracted from energy distribution curves.  
This occurs because of a combination of many-body effects and the 
presence of an energy gap, along with the associated coherence factors.
Analysis of such MDC dispersions can give important information on the
microscopics of high temperature superconductors.
\typeout{polish abstract}
\end{abstract}
\pacs{74.25.Jb, 74.72.Hs, 79.60.Bm}
\end{minipage}}

\maketitle
\narrowtext

Traditionally, practitioners in angle resolved photoemission spectroscopy 
(ARPES) have analyzed data at fixed momentum as a function of binding 
energy, so-called energy distribution curves (EDCs).
Recent advances in analyzer technology have allowed the probing of 
electronic states via ARPES to a
much higher precision in momentum space than previously 
attainable\cite{ADAM}.  This 
has led to the realization that additional information can be 
obtained by analyzing data at fixed binding energy as a function of 
momentum, so-called momentum distribution curves.  Such MDCs have been 
used in the high temperature cuprate superconductors for a variety of 
purposes, including the testing of the marginal Fermi liquid 
hypothesis\cite{VALLA1,VALLA2}, and the elucidation of a dispersion kink along 
the nodal direction\cite{KINK1}, the origin of which is currently being 
debated\cite{KINK2,KINK3,KINK4}.

Analysis of MDCs in the normal state, or in the superconducting state 
along the nodal direction, is relatively straightforward because of the 
absence of an energy gap\cite{KINK2}.  As we demonstrate in this paper, 
qualitative changes occur in the MDCs due to the energy gap.  By 
analyzing MDC dispersions, one can gain important information on 
many-body effects in the superconducting state.

The data reported in this paper were obtained at the Synchrotron 
Radiation Center, Wisconsin, using a Scienta SES 200 analyzer, and were 
previously used in earlier work\cite{KINK2}.  A photon energy of 22 eV was 
employed with the optimal doped ($T_c$=90K)
Bi$_2$Sr$_2$CaCu$_2$O$_{8+\delta}$ (Bi2212) sample in a $\Gamma-M$
polarization geometry.  The chemical potential was determined from a 
polycrystalline Au sample in electrical contact with the Bi2212 sample.

As previously reported\cite{KINK2}, we have taken data for a number of 
momentum cuts in the Brillouin zone both in the normal and 
superconducting states.  In this paper, we will concentrate our attention 
on a particular momentum cut intermediate between the $(\pi,\pi)$ direction 
where the superconducting gap vanishes and the $(\pi,0)$ region where the 
superconducting gap is maximal.  The reason for avoiding $(\pi,0)$ 
is that a combination of matrix element effects, 
superstructure images, and the pseudogap, complicate the 
interpretation of MDCs in this region of the zone\cite{HELEN} (for the
chosen cut, these complications are not present).
MDC and EDC dispersions were obtained from the maxima of the 
respective curves.

\begin{figure}
\epsfxsize=3.4in
\epsfbox{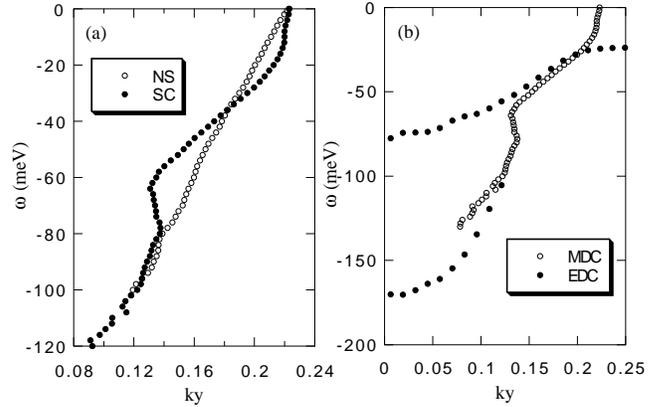}
\vspace{0.3cm}
\caption{(a) MDC dispersion in the superconducting state (SC, T=40K) versus that
in the normal state (NS, T=140K).
(b) MDC versus EDC dispersion in the superconducting state.
k$_y$ is in units of $\pi/a$.  For this momentum cut, k$_x$=0.59$\pi/a$.}
\label{fig1}
\end{figure}

In Fig.~1a, MDC dispersions are shown in both the normal and 
superconducting states.  The normal state dispersion is roughly linear in 
${\bf k}$ in the energy range of interest.  In 
the range of 20-60 meV, the superconducting dispersion is also linear, 
but with a slope approximately half that of the normal state, as noted 
earlier\cite{VALLA2}.  This implies an additional many-body renormalization of 
the superconducting state dispersion relative to that in the normal 
state.  Another effect of this renormalization can be seen at 
binding energies higher than 60 meV, where the dispersion goes almost
vertical before recovering back to the normal state dispersion.

To understand this effect in greater detail, we compare in Fig.~1b the 
dispersions in the superconducting state obtained from MDCs and EDCs.  As 
noted in an earlier paper\cite{KINK2}, the EDC dispersions 
contain two branches, a lower binding energy ``quasiparticle'' branch, and a 
higher binding energy branch (known as the ``hump'' in the $(\pi,0)$ 
region).  We see, then, that the vertical part of the MDC dispersion 
corresponds to a crossover between the low energy and high energy 
EDC branches.  These effects are typical of electrons interacting with a 
bosonic mode\cite{PARKS,ND}, and the mode in the current case has been 
identified as a spin exciton by some authors\cite{MODE,KINK2,KINK4} and a 
phonon by others\cite{KINK3}.
As can be seen from Fig.~1b and also noted above (Fig.~1a), the 
renormalization is an {\it additional} effect associated with the 
superconducting state, and thus unlikely to be due to a phonon.
Moreover, in Fig.~2, we show the MDC and EDC
dispersions calculated from the spin exciton model\cite{ESCH}.
We see that this calculation gives a good description of the experimental
data of Fig.~1b, and demonstrates the pronounced effect
of momentum dependent many-body interactions on the shape of the MDC 
and EDC dispersions.
This calculation also reproduces the difference between the MDC and EDC
dispersions in the ``linear" regime (20-60 meV),
which as we demonstrate below, does not occur in a BCS model.
The difference in the spin exciton case is associated with the
mode energy scale.

\begin{figure}
\epsfxsize=3.4in
\epsfbox{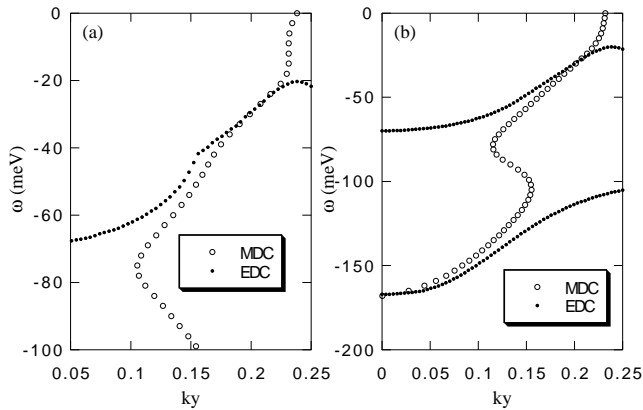}
\vspace{0.3cm}
\caption{MDC and EDC dispersions in the superconducting state from the
spin exciton model of Ref.~\protect\onlinecite{ESCH} ($\Gamma$=10 meV, mode
energy $\Omega$= 39 meV, coupling constant g=0.65 eV,
maximum gap $\Delta_0$=46 meV).
The energy resolution is $\sigma$=1 meV in (a) and 7 meV in (b).  In (a),
the weak kink in the EDC quasiparticle branch marks the mode energy, which
is washed out in (b) due to resolution.  The higher energy structure associated
with the S shape in the MDC dispersion is due to the strong frequency
dependence of the self-energy around $\Delta_0$+$\Omega$.}
\label{fig2}
\end{figure}

For the remainder of the paper, we concentrate on the low binding 
energy range, where the data are characterized by a renormalized 
``quasiparticle'' branch, and a simpler analysis of the data is possible.
Returning to Fig.~1a, we note that for binding 
energies lower than 20 meV, the MDC dispersion in the superconducting 
state goes almost vertical, and at zero energy is close to the
normal state Fermi 
momentum.  By looking at Fig.~1b, where the MDC and EDC dispersions are 
compared, we notice that the upturn in the MDC dispersion 
corresponds to entering the subgap region identified from the EDC dispersion.

Unfortunately, a model independent analysis of the data is somewhat 
impractical, as noted in passing in an earlier paper\cite{KINK2}.  This 
can be easily seen by a quick look at BCS theory.  In this theory, the 
effect of superconductivity can be represented by a self energy of the 
form $\Delta_{\bf k}^2/(\omega+\epsilon_{\bf k}+i0^+)$ where $\Delta_{\bf k}$ 
is the superconducting energy gap and $\epsilon_{\bf k}$ the normal state 
dispersion. Note that this self-energy has a 
non-trivial dependence on momentum (that is, it is not linear in ${\bf k}$), 
and thus invalidates a simple Lorentzian analysis of the MDCs.  

We have thus looked at a simple model to describe the MDC 
superconducting dispersion in the low energy range.  Simple BCS theory 
will not work, since by definition it has no solution in the subgap 
energy range.  The simplest generalization is to use a self-energy of the 
form\cite{PHEN} $-i\Gamma_1 + \Delta_{\bf k}^2/(\omega+\epsilon_{\bf k}
+i\Gamma_0)$.
The case where $\Gamma_1=\Gamma_0=\Gamma$ is just a broadened version of BCS 
theory\cite{SCHRIEFF} and we find that it gives a good account of the data.

For $\Delta_{\bf k}$, we assume a d-wave energy gap of the form 
$\Delta_0(\cos(k_xa)-\cos(k_ya))/2$ where $\Delta_0$ is fit by the 
energy of the quasiparticle peak in the EDC at the Fermi momentum.
We find that $\epsilon_{\bf k}$ is consistent with our earlier tight binding 
fit to normal state data\cite{NORM95} if a scaling factor, $z$, 
is introduced to account for the additional many-body renormalization of the 
superconducting state dispersion relative to the normal state one discussed 
above (for the momentum cut considered here, $z=0.61$).
The peak energy, 24 meV, sets $\Delta_0$ to be 46 meV.
At this stage, we will assume all the broadening is due to $\Gamma$, which is
obtained by fitting the top of the EDC peak at the Fermi momentum (giving
15 meV).  The effect of energy and momentum resolution broadening will be
treated later.

\begin{figure}
\epsfxsize=3.4in
\epsfbox{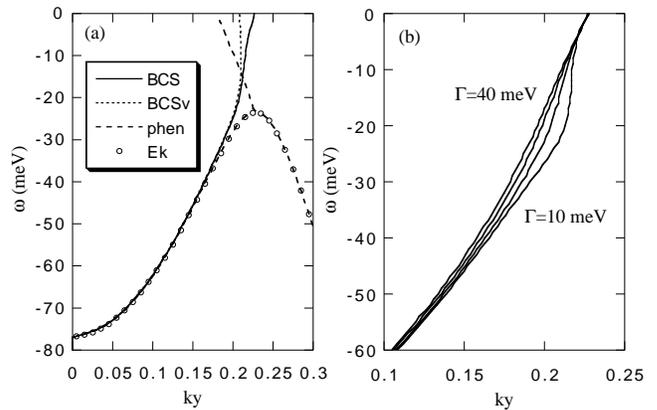}
\vspace{0.3cm}
\caption{Theoretical MDC dispersion in the superconducting state.  (a) Curves
correspond to Eq.~1 (BCS), second term in Eq.~1 only (BCSv), and Eq.~2 (phen).
The circles are the BCS energy dispersion, $\omega=-E_k$.  The parameters
are listed in the text ($\Gamma$=15 meV). (b) Results from Eq.~1 as a
function of the broadening parameter, $\Gamma$.}
\label{fig3}
\end{figure}

In Fig.~3a, we show our theoretical MDC dispersion, and compare it to 
some alternate theoretical curves to be discussed below.  To appreciate 
these results, we remind the reader that the broadened BCS spectral 
function can be written as\cite{SCHRIEFF}
\begin{equation}
\pi A({\bf k},\omega) = \frac{u_{\bf k}^2\Gamma}{\Gamma^2+(\omega-E_{\bf k})^2}
+ \frac{v_{\bf k}^2\Gamma}{\Gamma^2+(\omega+E_{\bf k})^2}
\end{equation}
where $u_{\bf k}$ and $v_{\bf k}$ are the BCS coherence factors, and
$E_{\bf k} = \sqrt{\epsilon_{\bf k}^2 + \Delta_{\bf k}^2}$ the BCS 
quasiparticle energies.  The simplest MDC is for $\omega=0$.  In this 
case, the right hand side simply reduces to $\Gamma/(\Gamma^2+E_{\bf k}^2)$.
Ignoring the weak variation of $\Delta_{\bf k}$ with ${\bf k}$, one has a peak
centered at $\epsilon_{\bf k}=0$, i.e., at the Fermi momentum, ${\bf k_F}$. 
An important point is that for this case, the coherence factors drop out,
and it is for this reason the peak is at ${\bf k_F}$.

Now consider $\omega < 0$ (occupied states), but within the subgap region.
If it were not for the coherence factors, the MDC peak would still be
centered at ${\bf k_F}$, i.e., the dispersion would be vertical.  But, the
coherence factors skew the peak to be centered at ${\bf k} < {\bf k_F}$.
This trend becomes more pronounced as $\Gamma$ increases, as can be seen in
Fig.~3b, where the MDC dispersion increasingly resembles the normal state.

As an exercise, we show the MDC dispersion in Fig.~3a, but where only the second
term in Eq.~1 is included.  In this case, the MDC dispersion becomes vertical
in the subgap region, but with an $\omega=0$ value significantly displaced
from ${\bf k_F}$, the displacement being due to the skewing caused by the
${\bf k}$ dependence of $v_{\bf k}$.  Physically, this behavior could occur if
the $\Gamma$ value for the first term in Eq.~1 was signficantly smaller than
that for the second term.  We note that several microscopic
theories for the cuprates do in fact predict this behavior\cite{LEE}.  That
is, the MDC dispersion is not only a sensitive test of the coherence factors,
but also particle-hole symmetry of the self-energy as well, even when looking
at just the occupied states.  Moreover, as we discuss below, energy resolution
will cause this skewing to occur as well.

Another curve is shown in Fig.~3a, and that is where $\Gamma_0$ is set to 0.
This self-energy is essentially the one used in our earlier work\cite{PHEN} in
the superconducting state, and corresponds to having no broadening in the
BCS part of the self-energy.  Although this model gives a good description of
the low energy part of the EDC at ${\bf k_F}$, it gives an erroneous MDC
dispersion.  This can be understood from the spectral function of this model
\begin{equation}
\pi A({\bf k},\omega) = \frac{\Gamma}{\Gamma^2+(\omega+E_{\bf k})^2
 [(\omega-E_{\bf k})/(\omega+\epsilon_{\bf k})]^2}
\end{equation}
It has been factored in such a way to emphasize the $\omega < 0$ branch.
Note this is not of the form of Eq.~1.  In particular, the ``coherence''
factors are $\omega$ dependent.  At $\omega=0$, the MDC is zero at ${\bf k_F}$,
that is the MDC has a minimum, rather than a maximum as in Eq.~1.
Because of this, the
MDC dispersion has a very strange behavior in the subgap region.  Moreover,
for $\omega < -\Delta$, the MDC dispersion becomes
similar to the EDC dispersion, reflecting the dispersion of $E_{\bf k}$.
This is because of the
peculiar ``coherence'' factors, which lead to a sharp second peak in the MDC
for ${\bf k} > {\bf k_F}$ corresponding to the particle-hole image
of the dispersion for positive $\omega$, as shown in Fig.~4a.  This subsidiary
peak is strongly reduced in Eq.~1 (consistent
with experiment), as can also be seen in Fig.~4a.  To test this further,
we have done two dimensional $\omega-{\bf k}$ intensity plots, and find that
Eq.~1 gives an intensity profile similar to experiment.  This is in
contrast with results obtained from Eq.~2, reminiscent of pure
BCS theory (i.e., zero broadening), where a pronounced intensity is seen
for ${\bf k} > {\bf k_F}$, reflecting the ``backbending''
expected from the $E_{\bf k}$ dispersion.  This is not seen in
experiment, and reiterates our point that the
MDCs and intensity profiles are quite sensitive to self-energy effects and
coherence factors.

\begin{figure}
\epsfxsize=3.4in
\epsfbox{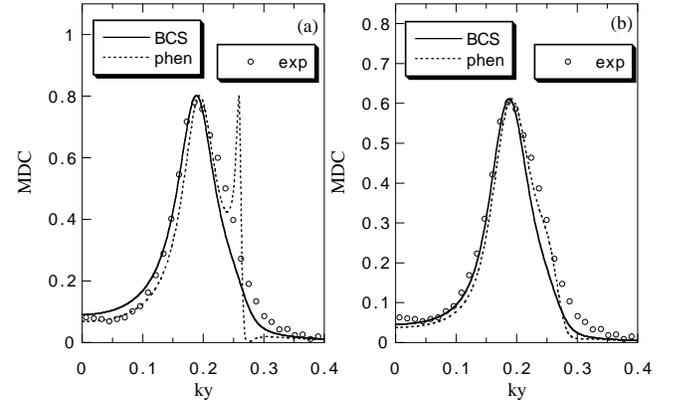}
\vspace{0.3cm}
\caption{(a) Theoretical MDC in the superconducting state from Eq.~1 (BCS)
and Eq.~2 (phen), compared to experiment (open circles).
$\omega$= -30 meV, and $\Gamma$=15 meV.
(b) Same results, but including energy resolution ($\sigma$=7 meV,
$\Gamma$=10 meV).}
\label{fig4}
\end{figure}

We now discuss the effects of momentum and energy resolution.  We have
verified by calculation that the small momentum window of the Scienta
analyzer (a
rectangle of dimensions 0.01$\pi/a$ along the cut, and 0.02$\pi/a$ transverse
to the cut) has no effect on the results.  This is not true for the energy
resolution.  The latter can be determined by fitting the leading edge of
the Au spectrum, which for the present data gives a Gaussian $\sigma$ of 7 meV
(FWHM of 16-17 meV).  As an initial exercise, let us assume that all the
broadening is due to energy resolution (for 30 meV FWHM, this would yield
a $\sigma$ of 12.8 meV).  Then, in the BCS case with the
spectral function as delta functions, the MDCs are very easy to
determine.  For an energy gap much larger than temperature (satisfied here,
since the EDC peak energy at the Fermi momentum is 24 meV, and the temperature
is 40K), the $u_{\bf k}$ term drops out ($f(E_{\bf k})$ is essentially zero).
Doing the energy resolution convolution, the ARPES intensity is simply
$v_{\bf k}^2e^{-(\omega+E_{\bf k})^2/2\sigma^2}$.  Since the intensity is
totally controled by the $v_{\bf k}$ term, the MDC dispersion will obviously
be skewed, as demonstrated in Fig.~5a, where this result is compared to the
previous broadened BCS case of Fig.~3a.  In particular, the MDC dispersion
at zero energy yields a momentum value which is slightly displaced from the
normal state Fermi momentum, and the MDC dispersion in the subgap region is
more vertical than in the $\Gamma$ broadened case.

\begin{figure}
\epsfxsize=3.4in
\epsfbox{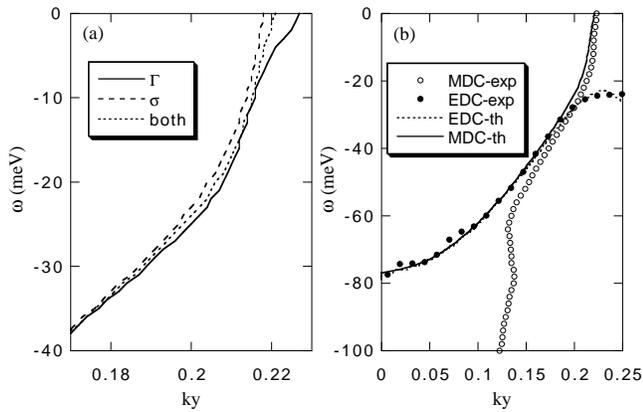}
\vspace{0.3cm}
\caption{(a) The effect of energy resolution on the MDC dispersion from Eq.~1.
The curves correspond to $\Gamma$=15 meV, $\sigma$=0 ($\Gamma$), $\Gamma$=0,
$\sigma$=12.8 meV ($\sigma$), and $\Gamma$=10 meV, $\sigma$=7 meV (both).
(b) Comparison of theoretical (th) and experimental (exp) MDC and EDC
dispersions. For theory, $\Gamma$=10 meV, $\sigma$=7 meV.}
\label{fig5}
\end{figure}

Also shown in Fig.~5a is the more realistic case where both effects are
incorporated.  Given the actual experimental $\sigma$ of 7 meV, a $\Gamma$
of 10 meV is necessary to reproduce the 30 meV FWHM of the EDC peak.
As expected, the combined result is intermediate between the two limiting
cases.  Moreover, the inclusion of energy resolution lessens the difference
between the MDC profiles of the two self-energy models presented in Fig.~4a,
as illustrated in Fig.~4b.  An advantage of the model of Eq.~2
is that it can account for the extra experimental weight on the trailing
(unoccupied) edge of the MDC peak not present in the broadened BCS
model.  In fact, by looking at experimental MDCs at higher binding energies,
a weak shoulder does develop on the trailing edge, corresponding to the
expected particle-hole image discussed earlier.  This image should become
better defined with improved resolution and statistics.

In fact, it is somewhat remarkable that the simple form of
Eq.~1 does such a good job in describing the data.  In Fig.~5b,
we compare
the MDC and EDC dispersions of Fig.~1b to our calculation of Fig.~5a.  Confining
ourselves to the ``quasiparticle'' branch, the agreement of experiment and
theory is quite good.  But, the discrepancy between
the MDC and EDC dispersions in the ``linear'' regime (20-60 meV) is not
reproduced by BCS
theory, though it can be accounted for in the spin exciton model as
demonstrated in Fig.~2.

In conclusion, we find that the MDCs and resulting dispersions are non-trivial
in the superconducting state, and give important information on the electron
self-energy and coherence factors.  We feel that a more detailed study of
MDCs, both in the superconducting and pseudogap phases, will give important
insights into the microscopics of high temperature cuprate superconductors.

This work was supported by the U. S. Dept. of Energy, Office of Science,
under contract W-31-109-ENG-38, and the National Science Foundation DMR 9974401.

\end{document}